\documentclass[
  reprint,
  aps,
  prl,
  amsmath,
  amssymb,
  superscriptaddress
]{revtex4-2}

\usepackage{graphicx}   
\usepackage{dcolumn}    
\usepackage{bm}         
\usepackage{siunitx}    
\usepackage{xcolor}     
\usepackage{hyperref}   
\usepackage{subcaption} 
\usepackage{booktabs}

\begin{document}

\title{%
Quantifying the Features of an Amorphous Solid's Local Yield Surface%
}

\author{Spencer Fajardo}
\affiliation{Materials Science and Engineering, Johns Hopkins University, Baltimore, MD USA}

\author{Paul Desmarchelier}
\affiliation{ Sorbonne Universit\'e, CNRS, Physico-chimie des \'Electrolytes et Nanosyst\`emes Interfaciaux, PHENIX, 75005 Paris, France}

\author{Sylvain Patinet}
\affiliation{PMMH, CNRS, ESPCI Paris, Universit\'e PSL, Sorbonne Universit\'e, Universit\'e Paris Cit\'e FRANCE}

\author{Michael L. Falk}
\affiliation{Materials Science and Engineering, Johns Hopkins University, Baltimore, MD USA}
\affiliation{Mechanical Engineering, Johns Hopkins University, Baltimore, MD USA}
\affiliation{Physics and Astronomy, Johns Hopkins University, Baltimore, MD USA}
\affiliation{Hopkins Extreme Materials Institute, Johns Hopkins University, Baltimore, MD USA}
\affiliation{Data Science and AI Institute, Johns Hopkins University, Baltimore, MD USA}

\date{\today}

\begin{abstract}

In two-dimensional Lennard–Jones glasses, mechanical probing reveals that local yield surfaces are dominated by regions with a positive second derivative of the yield stress with respect to the loading angle. Each feature corresponds to a shear transformation zone and a characteristic non-affine displacement field at yield. Most features are well described by a combined Schmid–Mohr–Coulomb criterion parameterized by a weak-plane orientation, a critical stress, and a pressure sensitivity. The resulting parameter statistics clarify how the onset of plastic flow is governed by the population of discrete yielding features encoded in the amorphous structure.

\end{abstract}

\maketitle

Amorphous solids are a broad class of materials that exhibit non-crystalline solid structure \cite{JohnsonBulkTechnology} often produced by rapidly cooling a liquid or depositing a vapor so as to inhibit the kinetics of crystallization \cite{Klement1960187869b0}. Understanding the mechanical behavior of these materials beyond the elastic regime remains hindered by an inability to rationalize plastic response in terms of distinct features of the amorphous structure \cite{Schuh2007MechanicalAlloys, Eckert2007MechanicalComposites, Nicolas2018DeformationModels, Rodney2011ModelingScale, Argon1979PlasticGlasses, Spaepen1977, Hufnagel2016DeformationExperiments.}. Plasticity in amorphous solids is thought to be mediated by shear transformation zones (STZs) \cite{Falk1998}, discrete local structural features that are susceptible to rearrangement under shear and act as flow-defects. An important question that remains is how these STZs embedded in the material's structure control the yield response at the nanoscale. In this work, we link the nanoscale structure of a metallic glass to its plastic behavior by adapting continuum-mechanical concepts to the nanoscale.

\par In continuum mechanics, the concept of a yield surface is commonly used to describe the yielding conditions of a material under complex loading conditions\cite{ phillips1965,michno1976historical,Schuh2003,hill1998mathematical}. The yield surface specifies the stress conditions that trigger plastic flow, defining  an envelope within which the material behaves elastically. We extend this idea from continuum mechanics to the nanoscale to establish a direct connection between the yield stress concept and the atomic-scale physics of the STZs that control plasticity.

\par Here we show that most positive-curvature segments of the local yield surface can be mapped one-to-one to STZ-like rearrangements and fitted by a Schmid–Mohr–Coulomb form, enabling parameter statistics vs quench rate. We describe the system we use as a model of amorphous solids and the nanoscale mechanical probing technique used to induce plasticity. By deploying this technique, we measure stress at the point of plastic onset to determine the angular and pressure dependence of these plastic events. Finally, we show that the yielding behavior of the model amorphous solid can be described by a yield criterion that incorporates aspects of the Schmid and Mohr-Coulomb yield criteria. This form is used to collect statistics on physical parameters that characterize the material's yielding behavior as a function of the preparation protocol. These results show, from a unitary flow-defect perspective, that yield stress and pressure sensitivity increase as glass is quenched more deeply.

\begin{figure}[t]
\begin{minipage}[t]{1\columnwidth}
\vspace{0pt}
\includegraphics[width=\linewidth]{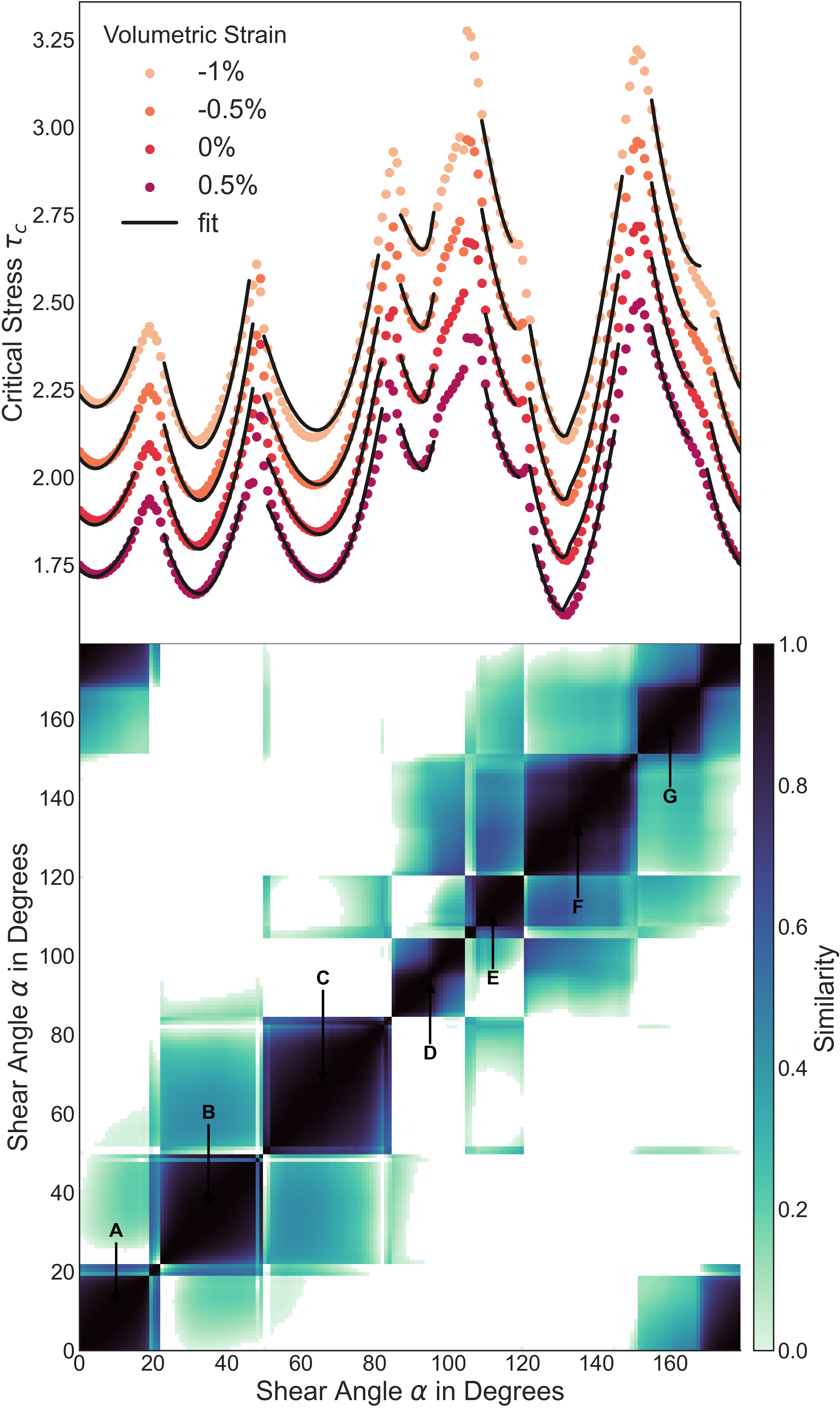}
\end{minipage}
\hfill
\caption{%
Top: Critical stress as a function of shear angle, defining a local yield surface. Symbols and lines correspond, respectively, to the atomistic data and their fit (Eq.~\ref{Eq:STZsegment}); colors indicate the volumetric-strain conditions. Bottom: Similarity metric (Eq.~\ref{eq:dispcomparison}) quantifying the alignment of plastic non-affine displacement fields; regions along the diagonal are labeled A--G.}
\label{fig:combined}
\end{figure}

\par \textit{Methods}--- 
We study the local plastic response of a simulated two-dimensional, binary Lennard-Jones (LJ) glass-forming system \cite{Lancon,Leoni,Barbot2018}. Glass structures are generated using the procedure described in Ref. \cite{Shi2005StrainSolids}. The LJ pair potential cutoff with energy and force shifting is adapted from Wahnström \cite{Wahnstrm1991}; details of the smoothing and shifting procedure are provided in the Supplemental Material \cite{SM}. 

\par Each system consists of $15,625$ atoms composed of two different species, designated as Large (L) and Small (S), where the number of large atoms ($N_L$) and the number of small atoms ($N_S$) is given by the ratio $N_L:N_S = (1+\sqrt(5)):4$, both with mass $m=1$. All quantities are reported in reduced Lennard–Jones units: length is measured in units of $a_{SL}$ the distance at which an S and an L atom have zero potential energy; energy is measured in units of the bond energy between S and L atoms, $\epsilon_{SL}$; and time is measured in units of $\tau = a_{SL}\sqrt{m/\epsilon_{SL}}$. The zero-potential-energy distances between like atom types are $a_{LL}=2\sin(\pi/5)a_{SL}$ and $a_{SS}=2\sin(\pi/10)a_{SL}$ respectively. The bond energies between two like atoms are $\epsilon_{LL} = \epsilon_{SS} = 0.5\epsilon_{SL}$.  In this system, the glass transition temperature is known to be approximately $T_{g} \approx 0.325 \epsilon_{SL}/k_B$ \cite{Lancon}. After initializing the atoms with an initial density of $1/V_{atom}$, where $V_{atom} = 0.976 a_{SL}^2$ and imposing periodic boundary conditions, the system is heated to a melting temperature $T_m = 2T_g$ using a Nose-Hoover thermostat and equilibrates at this temperature before it is cooled down to a final temperature $T_f = 0.092T_g$. Slower quenching rates result in glasses that are more deeply quenched with lower potential energies that are typically harder and more brittle than more quickly quenched glasses  \cite{Hufnagel2016DeformationExperiments.,Eckert2007MechanicalComposites,JohnsonBulkTechnology}. We study glasses prepared at four cooling rates: $7.632\times10^{-5}T_g/\tau$, $7.632\times10^{-6}T_g/\tau$, $7.632\times10^{-7}T_g/\tau$, and $8.885\times10^{-8}T_g/\tau$.  We will refer to these glass preparation methods as G1, G2, G3, and G4, where G1 is the fastest cooling rate, and G4 is the slowest.

\par The local yield stress (LYS) method systematically measures the shear stress at which a local region of a material yields when subjected to incrementally increasing shear strains \cite{Patinet2016ConnectingSolids,Ruan2022,Richard2020PredictingIndicators}. Locations in a glass structure with low LYS have been shown to correspond to sites of plasticity upon remote loading \cite{Patinet2016ConnectingSolids,Richard2020PredictingIndicators}.
Here, we employ the LYS procedure to calculate the critical stress $\tau_{c}$, which is the shear stress in the loading direction at the point of yielding, by applying athermal quasi-static shear (AQS) \cite{Maloney, Malandro} to circular regions of the system 10$a_{SL}$ in radius. Two different subregions are defined within this circle: an outer annular region from 5$a_{SL}$ to 10$a_{SL}$ within which the atoms will be constrained to shear, and an inner region from the center to 5$a_{SL}$ free to mechanically equilibrate as the outer region is deformed. The sample is affinely deformed in pure shear strain increments of $\Delta\gamma = 10^{-4}$. Between each strain step, the positions of the atoms in the inside region are relaxed to a minimum energy configuration via a conjugate gradient scheme. This process is repeated until the stress in the loading direction decreases from one step to the next, signaling plastic deformation. At this point, the two-dimensional stress state at the onset of instability $\bar{\bar{\sigma}}^{c}$ is recorded. Finally, the critical shear stress is computed as 
\begin{equation} \label{eq:2}
\tau_{c} = (\bar{\bar{\sigma}}_{oi} : \hat{S}_{ij}),
\end{equation}
where $\hat{S}_{ij}$ is a unit tensor in the direction in which the shear strain is applied.

\par \textit{Results and Analysis}--- 
We use the above-described method to construct yield surfaces of various circular regions from our system. This is done by probing the critical shear stress over a range of angles, thereby mapping out the maximum elastic stress attainable before the onset of plasticity. Figure \ref{fig:combined} shows a plot of $\tau_c$ as a function of shear angle $\alpha$ for a particular region of our glass.
We observe that the yield surface shown in Figure \ref{fig:combined} is composed of wells, as one would expect in materials governed by a yield criterion wherein slip occurs along weak planes along particular shear directions.

To confirm the hypothesis that the yield surface can be decomposed into distinct regions corresponding to the activation of particular STZs with characteristic shear directions, we divide the yield surface into regions of positive and negative curvature. Positive curvature segments constitute $69.25\%$ of all yield surfaces. We then seek to associate a distinct STZ with each of these regions. We hypothesize that the negative curvature cusps correspond to transitions between activating one distinct STZ and another as the angle of applied strain is varied.

\par Previous work has shown that the yielding response in models of metallic glasses is pressure-dependent, consistent with the Mohr-Coulomb yield criterion \cite{Lund2003,Lund2004,Zhao2016,Schuh2003}.
To describe the orientation and pressure dependence of the plastic response due to the activation of differently oriented STZs with a variety of critical stresses, Lerbinger \cite{lerbinger:tel-03151944} developed a two-dimensional yield criterion that incorporates elements of both the Mohr-Coulomb and Schmid law of the form
\begin{equation}
\tau_c(\alpha) = \frac{\tau_0 - \phi p^{c}}{\cos(2(\theta-\alpha))}-\frac{\sigma_{yy}^c-\sigma_{xx}^c}{2}\tan(2(\theta-\alpha)).
\label{Eq:STZsegment}
\end{equation}
This criterion is meant to provide a piecewise continuous description of the portion of the yield surface associated with a single STZ. Here, the critical stress is a function of the shear angle $\alpha$ and depends on the orientation of the STZ, $\theta$, the intrinsic critical stress, $\tau_0$, which is the minimum shear required to activate the STZ in the absence of pressure, and the pressure dependence of the plastic response, $\phi$. In two dimensions, the hydrostatic pressure at the point of yielding is $p^{c} = -(\sigma_{xx}^c+\sigma_{yy}^c)/2$. The pressure dependence $\phi$ is typically negative, indicating that the critical stress increases under pressure \cite{Lund2003,Lund2004}. An example of this behavior is observed in data from differently strained samples as shown in Figure \ref{fig:combined}. For each sample, we generate additional pressure conditions by applying the same incremental shear process as in the LYS method, but with an applied hydrostatic strain. The second term on the right-hand side of Eq.~\ref{Eq:STZsegment} takes into account the anisotropic response of the material to stress, where $\sigma^c_{yy}-\sigma^c_{xx}$ represents the orthogonal shear stress projected along $\theta$ at the point of yielding. 

\par To determine the parameters $\theta$, $\tau_0$, and $\phi$, we associate positive-curvature regions spanning overlapping angular ranges across multiple volumetric strain conditions. Regions observed under only a single pressure state are excluded, since the yield-criterion parameters cannot be uniquely identified from a single condition. Regions spanning less than $5^\circ$ of the yield surface are also excluded.
Parameters are obtained by nonlinear least-squares fitting of Eq.~\ref{Eq:STZsegment} \cite{Vugrin2007}. Figure~\ref{fig:combined} shows the resulting fits. We find that $92.65\%$ of fitted points exhibit deviations of less than $5\%$ from their measured values.

\begin{figure}[t]
\centering
\includegraphics[width=1\columnwidth]{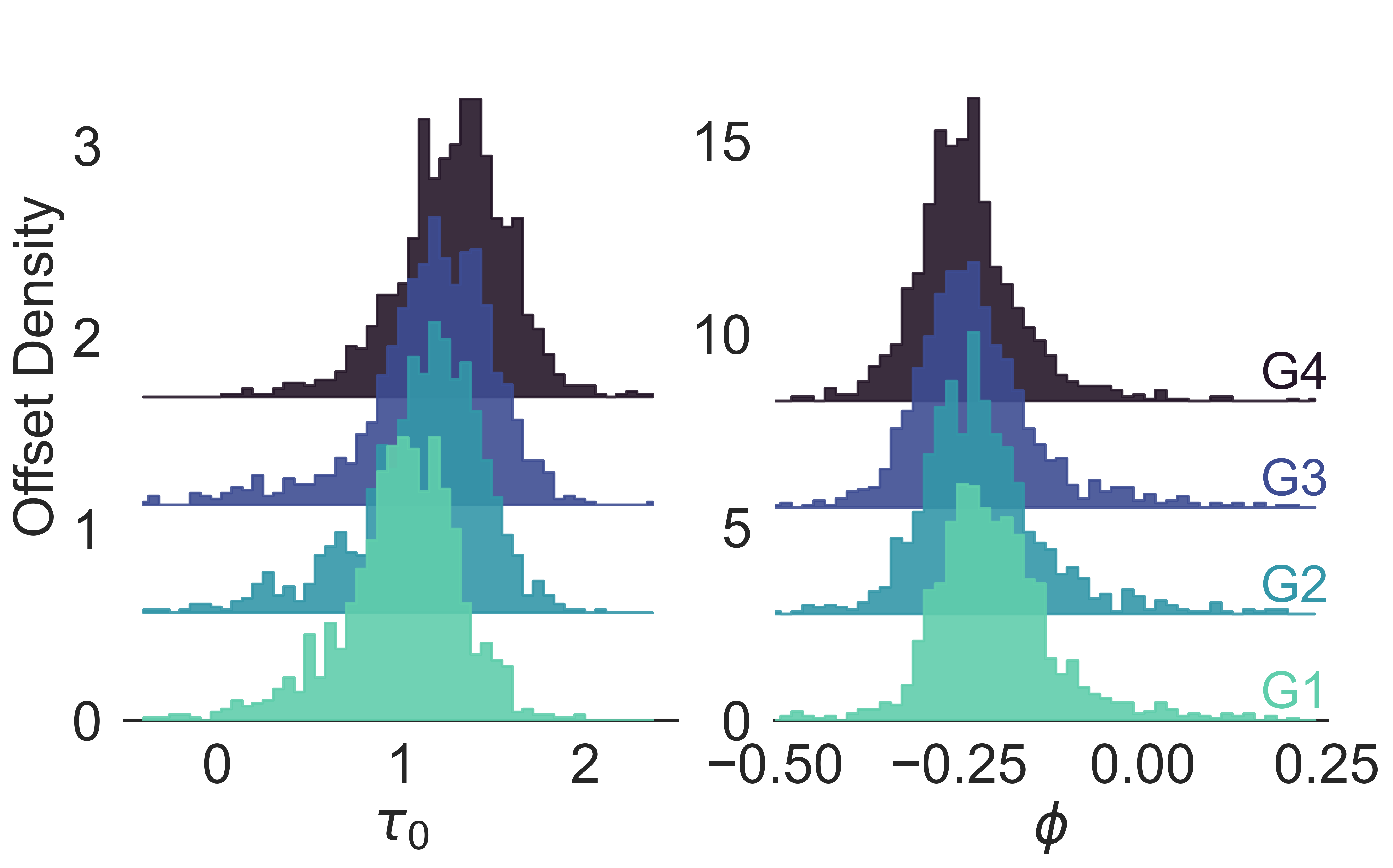} \\
\caption{Distributions of the intrinsic critical stresses $\tau_0$ and pressure dependencies, $\phi$, collected from $4$ different glasses. The values are colored by quench rate. Each distribution is shifted for clarity. The mean value of $\tau_0$ shifts higher as the cooling rate decreases, while $\phi$ shows a shift towards lower values at slower cooling rates. See Table \ref{tab:params} for more detail. Quench rate dependencies of the mean values of $\tau_0$ and $\phi$ are provided in Figure S1 of the Supplementary Materials.}
\label{fig:MCstatistics}
\end{figure}
By sampling $144$ independent regions extracted from each of four glass models, we collected statistics on $4726$ positive curvature regions as shown in Figure~\ref{fig:MCstatistics} and in Table \ref{tab:params}. 
This represents an average of 8.2 STZs per region accounting for between $82-84\%$ of the positive curvature regions of all samples. For more details on the percentage of yield surfaces fit, and the average error in fits see Supplemental Materials.

We analyze the cumulative distributions for both $\tau_0$ and $\phi$ excluding outliers. Statistical tests show that the distributions are consistent with normality.
We observe a systematic trend that glasses produced at lower cooling rates exhibit a 27.8\% higher average $\tau_0$, confirming our expectations \cite{Ozawa2018, Ozawa2020, Barbot2020,Rottler2005}. The average $\phi$ shifts 17.3\% towards more negative values as the cooling rate decreases, indicating that the pressure dependence of the yield stress increases in magnitude as the quench rate decreases. Both of these parameters exhibit a logarithmic relationship to the quench rate, in line with the literature (See Supplemental Material). \cite{Ozawa2018, Ozawa2020,Barbot2020,Rottler2005,Yu2007,Yu2010,Dubach2009}

\begin{table}[t]
\centering
\caption{Summary statistics for the intrinsic critical stress $\tau_0$ and pressure sensitivity $\phi$ across preparation conditions.
Cooling rates are provided in the methods section.}
\label{tab:params}

\setlength{\tabcolsep}{4pt} 
\begin{ruledtabular}
\begin{tabular}{lcccc}
Condition & $\langle\tau_0\rangle$ & s.d. & $\langle\phi\rangle$ & s.d. \\
\hline
G1: & $1.023 \pm 0.009$ & 0.337 & $-0.196 \pm 0.003$ & 0.105 \\
G2: & $1.141 \pm 0.011$ & 0.372 & $-0.211 \pm 0.004$ & 0.121 \\
G3: & $1.192 \pm 0.011$ & 0.357 & $-0.221 \pm 0.003$ & 0.091 \\
G4: & $1.307 \pm 0.009$ & 0.317 & $-0.230 \pm 0.002$ & 0.082 \\
\end{tabular}
\end{ruledtabular}
\end{table}

\begin{figure*}[t]
\centering
\includegraphics[width=1\textwidth]{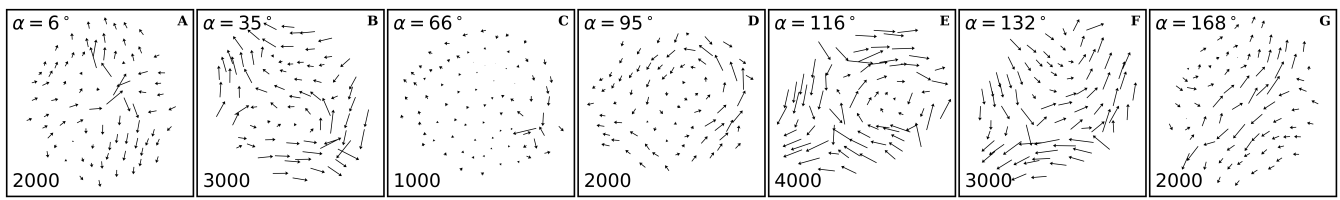} \\
\caption{Plastic displacement fields shown for the regions labeled A-G in Fig. \ref{fig:combined}(bottom), with their scaling factor in the bottom left. The fields can be easily clustered by direction through similarity using Eq. \ref{eq:dispcomparison}. From Fig. \ref{fig:combined}(top), each distinct deformation mode can be consistently associated with positive curvature regions throughout its angular domain.}
\label{fig:DispFields}
\end{figure*}

\par We next test the hypothesis that the positive curvature regions that we find to be well-characterized by Eq.~\ref{Eq:STZsegment} correspond to distinct STZs. Displacement fields are compared to characterize the plastic rearrangements that occur when a sample yields under applied shear. We compute plastic displacement by subtracting the positions of the atoms immediately before and after yielding. In doing so, we subtract the affine displacement imposed due to the shear at the boundary. In Figure \ref{fig:DispFields}, seven displacement fields are shown, each corresponding to a fitted minimum. The displacement fields are labelled A-G corresponding to the minima in the bottom half of Figure \ref{fig:combined}.
\par Notably, each STZ so identified has a quantifiably similar displacement field throughout its range. For a sample with $N$ atoms, the resulting normalized plastic displacement field is a $2N$ vector $\Vec{V_i}$ comprised of the x and y displacements for each atom. The inner product of any two such displacement fields quantifies the degree of similarity which can be expressed as a real number $r \in [0,1]$.
\begin{equation}
r = \frac{\Vec{V_1}\cdot\Vec{V_2}}{|\Vec{V_1}||\Vec{V_2}|}
\label{eq:dispcomparison}
\end{equation}
A value of 1 would signify that the two deformation responses were identical, whereas a value of 0 would indicate that the two responses were orthogonal. 

\par The lower panel of Figure~\ref{fig:combined} applies the metric provided in Eq.~\ref{eq:dispcomparison} to compare the plastic rearrangements triggered by applying shear at different angles to one sampled region. The positive curvature regions in the upper portion of the figure correspond closely with the dark regions in the lower plot. These dark regions denote similarity between the plastic rearrangements triggered at different loading angles.  The absence of dark regions off the diagonal suggests that each positive curvature region has a characteristic deformation response not found at any other angle of shear.

\par To check the consistency of this result among all fitted positive curvature regions in all samples, we compute the median value of similarity within each such region. Our dataset consists of $20,803$ positive-curvature regions across all $4$ glasses, previously identified as candidates for fitting. To test our hypothesis without bias from grouping, we consider each candidate region separately. We found statistically significant differences in the variance of the similarity metric across cooling procedures. Thus, we do not group across cooling procedures to assess the similarity. We find that $96.64\%$ of all STZs have a median similarity measure of $90\%$ or higher. This indicates that regions of positive curvature correspond with STZs that undergo markedly similar rearrangements. Panels A—G in Figure \ref{fig:DispFields} show displacement fields at the angle $\alpha$ corresponding to the closest direction to the fitted weak plane direction $\theta$ of each STZ. These displacement fields illustrate that distinct STZs exhibit characteristic displacement signatures as quantified using Eq.~\ref{eq:dispcomparison}

\par \textit{Discussion}--- 
In this study, we address two questions central to linking the STZ concept in amorphous solids to nanoscale plasticity. First, we test whether the nanoscale yield surface can be interpreted as the manifestation of discrete STZ-like entities. Our results show that this holds for a substantial fraction of the local yield surface in a two-dimensional Lennard–Jones glass. Second, we examine whether the associated plastic rearrangements uniquely characterize the corresponding portions of the yield surface. The data support this connection in the model system considered. We further find that the local yield surfaces display yielding signatures consistent with a combination of Mohr–Coulomb and Schmid-type criteria. This enables a decomposition of the yield surface into a discrete set of segments, each apparently associated with a single STZ. From this decomposition, we can estimate the STZ number density in the glass and quantify how their response depends on both shear orientation and pressure. These findings open a promising route for future investigations that aim to connect atomistic rearrangements to population-based predictive descriptions of glass microstructural evolution during plasticity in amorphous solids. 

\par \textit{Acknowledgements}--- 
This material is based upon work supported by the U.S. National Science Foundation (DMREF-2323718). This work was carried out at the Advanced Research Computing at Hopkins (ARCH) core facility (rockfish.jhu.edu), which is also supported by the U.S. National Science Foundation (OAC-1920103).

 \bibliography{references}

\clearpage
\onecolumngrid

\begin{center}
\textbf{\large Supplemental Material}
\end{center}

\renewcommand{\thefigure}{S\arabic{figure}}
\renewcommand{\thetable}{S\arabic{table}}
\setcounter{figure}{0}
\setcounter{table}{0}


\title{Supplemental Material for ``Quantifying the Features of an Amorphous Solid's Local Yield Surface''}

\maketitle

\onecolumngrid

\section*{S1. Smoothed Wahnström Potential (WAHNS)}

We use a modified version of the Wahnström binary Lennard--Jones (LJ) glass-forming system \cite{Wahnstrm1991, Shi2005StrainSolids}. Glass structures are generated following Ref.~\cite{Shi2005StrainSolids,Lancon,Barbot2018}. The pair potential is given by
\begin{equation}
E(r) =
\begin{cases}
4\epsilon\!\left[\left(\dfrac{\sigma}{r}\right)^{12} -
\left(\dfrac{\sigma}{r}\right)^{6}\right] - E_{0}, & r < r_{\mathrm{in}}, \\
C_0 - C_1(r-r_{\mathrm{in}}) - \dfrac{C_2}{2}(r-r_{\mathrm{in}})^2
 - \dfrac{C_3}{3}(r-r_{\mathrm{in}})^3 - \dfrac{C_4}{4}(r-r_{\mathrm{in}})^4,
 & r_{\mathrm{in}} < r < r_c, \\
0, & r > r_c .
\end{cases}
\end{equation}

Here $r$ is the interparticle distance, $\sigma$ is the LJ length scale, and $\epsilon$ is the bond energy. The interaction is truncated at $r_c = 2.5\sigma$ with a smooth polynomial continuation between $r_{\mathrm{in}} = 2\sigma$ and $r_c$ such that the potential and its first two derivatives are continuous. The coefficients $E_0$ and $C_i$ are fixed accordingly; full expressions are provided in the Supplemental Material of Ref.~\cite{Leoni}.

\section*{S2. Logarithmic Dependence of Fitted Parameters on Cooling Rate}

Figure~S1 shows the mean values of the fitted parameters $\tau_0$ and $\phi$ as functions of the cooling rate. In both cases, the data are well described by a logarithmic dependence, with 95\% confidence intervals indicated. This behavior is consistent with prior two- and three-dimensional simulations and experimental observations \cite{Ozawa2018,Ozawa2020,Barbot2020,Rottler2005,Yu2007,Yu2010,Dubach2009}.

\begin{figure}[b!]
\centering
\includegraphics[width=0.48\linewidth]{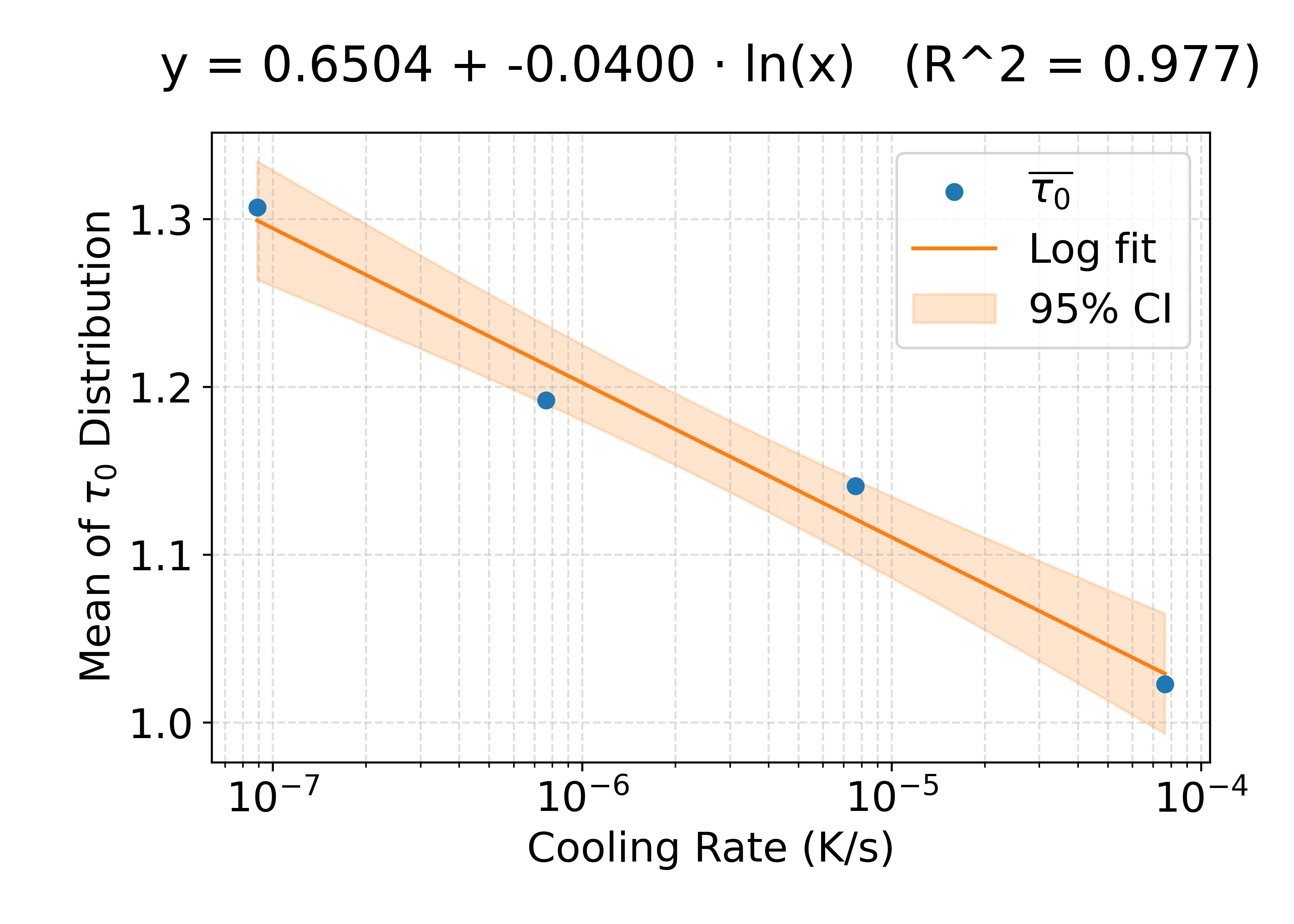}\hfill
\includegraphics[width=0.48\linewidth]{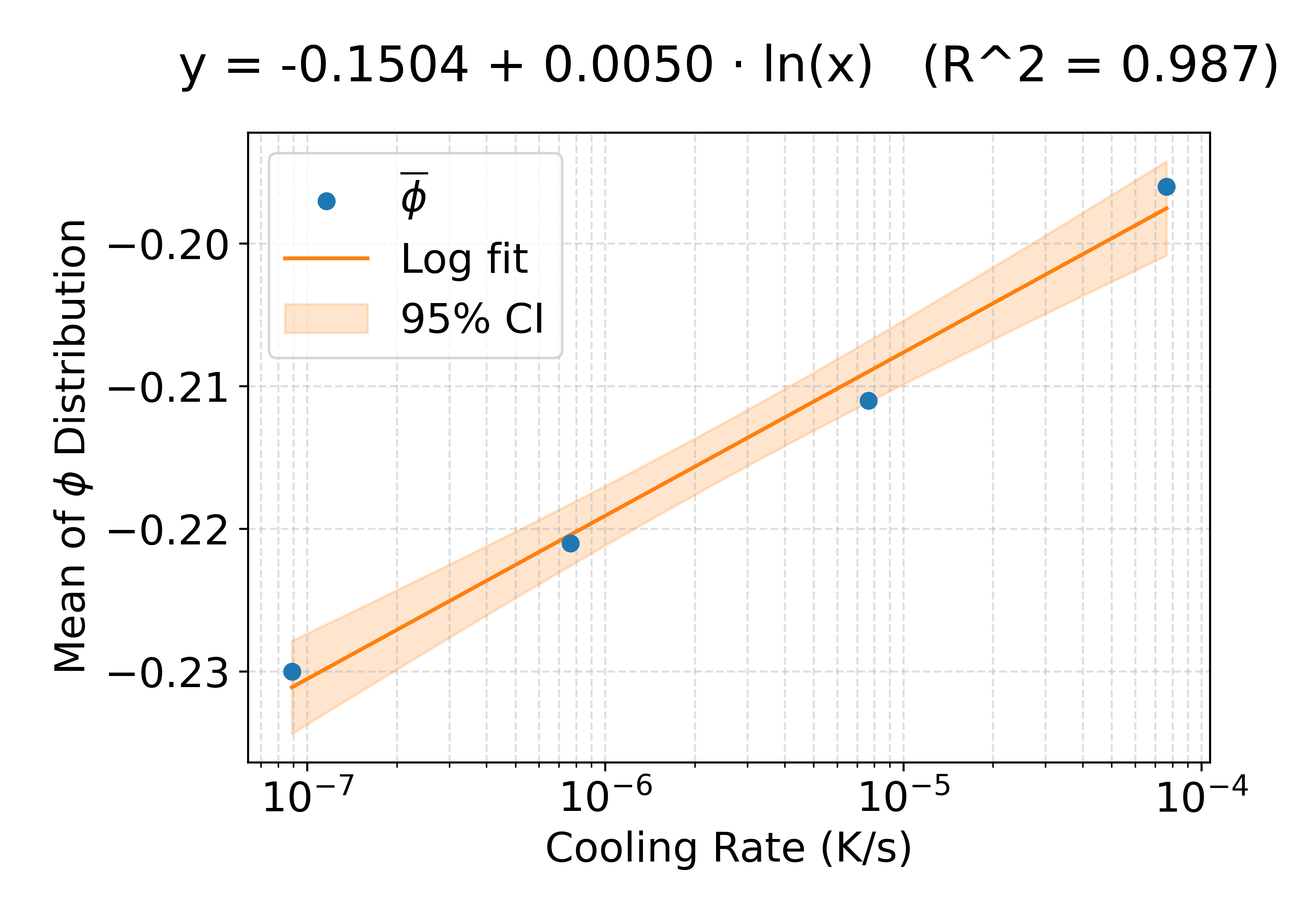}
\caption{(a) Mean intrinsic critical stress $\tau_0$ versus cooling rate.  
(b) Mean pressure sensitivity $\phi$ versus cooling rate. Shaded regions indicate 95\% confidence intervals.}
\label{fig:S1}
\end{figure}

\section*{S3. Fitting Statistics Across Cooling Rates}

Table~S1 summarizes the fraction of the yield surface exhibiting positive curvature, the fraction retained for fitting, the total fitted portion of the yield surface, and the average percent error as functions of cooling rate.

\begin{table}
\centering
\caption{Fitting statistics across preparation conditions.}
\label{tab:S1}
\begin{tabular}{ccccc}
\toprule
Preparation protocol \qquad \qquad & Convex (\% of total) \qquad \qquad & Kept (\% of convex) \qquad \qquad & Fit (\% of total) \qquad \qquad & Avg. rel. error (\%) \\
\midrule
G1 & 69.6 & 82.1 & 57.2 & 2.75 \\
G2 & 69.5 & 84.2 & 58.6 & 4.93 \\
G3 & 68.8 & 84.1 & 57.9 & 2.39 \\
G4 & 69.0 & 82.8 & 57.1 & 2.91 \\
\bottomrule
\end{tabular}
\end{table}


\end{document}